\documentclass[acmsmall,screen,nonacm]{acmart}

\usepackage{amsmath}
\usepackage{textcomp}
\usepackage{tabularx}
\usepackage{multirow}
\usepackage{soul}
\usepackage{enumitem}
\usepackage{tcolorbox}
\newcommand{\figsharelink}{\url{https://figshare.com/s/bcdd07ff46760fd254ec}}
\newcommand{\inlineheadingbf}[1]{\medskip\noindent{\bfseries #1.}}
\newcommand{\inlineheadingit}[1]{\medskip\noindent{\it #1.}}

\newcommand{\llbox}[1]{%
  \begin{tcolorbox}[
    width=\columnwidth,
    colback=black!6,
    colframe=black!6,
    boxrule=0pt,
    arc=2pt,
    top=1.5mm, bottom=1.5mm, left=2mm, right=1.5mm
  ]
  #1
  \end{tcolorbox}%
}

\begin{document}

\title{On the Risks of using LLM-Generated Tests for Regression Testing}

\author{Mohammadali Charoosaei}
\email{mohammadali.charoosaei@uni.lu}
\affiliation{%
  \institution{SnT, University of Luxembourg}
  \country{Luxembourg}}

\author{Cedric Richter}
\email{cedric.richter@uni.lu}
\affiliation{%
  \institution{SnT, University of Luxembourg}
  \country{Luxembourg}}

\author{Mike Papadakis}
\email{michail.papadakis@uni.lu}
\affiliation{%
  \institution{SnT, University of Luxembourg}
  \country{Luxembourg}}

\begin{abstract}
Software is under constant evolution. Developers continuously add features, fix bugs, and refactor code, and any of these changes may inadvertently break existing functionality. Regression testing guards against such unintended effects by capturing expected behavior in test cases: when a test fails after a change, it signals that previously established behavior has been altered. LLM-based test generation aims to automate this process by generating regression tests directly from the code under test. While this approach can be beneficial when the implementation is correct, it becomes particularly problematic when the code contains faults. In such cases, generated tests may encode and preserve incorrect behavior, allowing faults to persist and potentially influence subsequent stages of software evolution. To investigate this risk, we apply LLM-based regression test generation to pull requests merged into the main branch of software projects and study the impact of the generated tests on subsequent project evolution. We distinguish between \emph{fault-revealing}, which assert correctly implemented behavior, and \emph{fault-enforcing tests}, which assert faulty behavior. We then empirically quantify the extent to which fault-enforcing tests can preserve bugs and hinder their correction during software evolution. Across 145 pull requests from SciPy, Qiskit, and pandas, 8\%--17\% of the generated tests are fault-enforcing, while only 2.4\%--4.8\% reveal faults. Most of the remaining tests do not provide any relevant signal, i.e., they pass on both versions or fail on both. Fault-enforcing tests persist over time: after several subsequent commits, 83\%--91\% of them are still relevant and pass. We also find that fault-enforcing tests may also accumulate over time: when the faults of all pull requests are combined in one codebase, 83\%--92\% remain enforced at the end of the commit history, and the developer-written test suite detects only 14\%--30\% of them. Taken together, our results reveal a fundamental risk of LLM-generated regression tests: without manual validation, they may encode faulty behavior as expected behavior, allowing bugs to persist across software revisions and largely evade detection by developer-maintained test suites. As a result, LLM-based regression testing can give rise to a new form of technical debt, in which accumulated fault-enforcing tests perpetuate defects and constrain software evolution.

\end{abstract}



\maketitle

\section{Introduction}
\label{sec:intro}

Modern software systems evolve continuously, with developers frequently submitting pull requests (PRs) that add features or fix bugs. Each change risks introducing a \emph{regression}: functionality that worked before breaks after the change. Regression testing~\cite{yoo2012regression} detects such regressions, so that developers can react and fix the broken behavior if necessary. However, writing high-quality regression tests is among the most expensive activities in software maintenance~\cite{leung1989insights,yoo2012regression} and requires substantial effort and domain knowledge. As a result, newly introduced behavior is often under-tested, leaving the project vulnerable to future regressions.

Recent advances in LLMs promise to reduce this effort. By exploiting code context, commit messages, and code diffs, LLM-based approaches can automatically generate executable regression tests for software changes. ChaCo~\cite{zhou2026chaco} augments tests at the pull-request level by targeting previously uncovered lines; Liu~et~al.~\cite{liu2025regression} evaluate LLM-generated regression tests for commits; SymPrompt~\cite{ryan2024symprompt} uses code-aware prompting for coverage-guided generation; and CoverUp~\cite{pizzorno2024coverup} iteratively extends test suites using coverage feedback. These approaches report substantial coverage improvements with little manual intervention, which suggests that LLMs could considerably reduce the effort of maintaining evolving test suites.

\begin{figure}[t]
\centering
\vspace{-0.5em}
\includegraphics[width=0.95\textwidth]{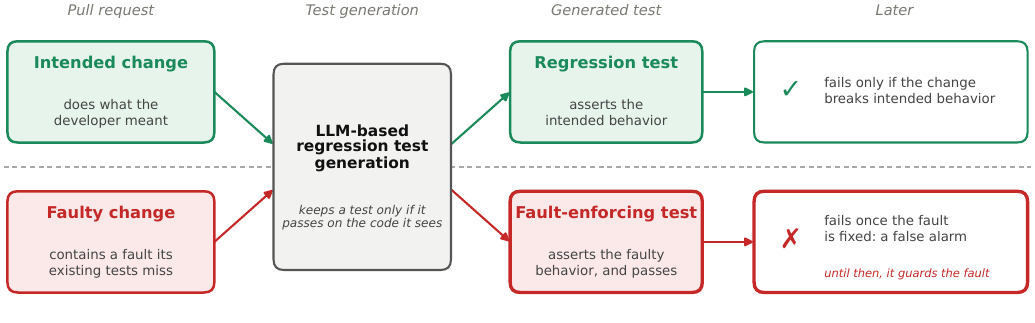}
\caption{LLM-based regression testings works under the assumption that a PR reflects the intended change. In a realistic setting where a PR might be \emph{faulty}, LLM-based regression testing risks to enforce the faulty behavior of the PR. We study how frequent LLMs enforce faulty behavior and how fault-enforcing tests propagate through the history of the codebase.}
\label{fig:motivation}
\vspace{-0.1em}
\end{figure}

Despite these efforts, LLM-generated regression tests differ fundamentally from a developer's testing process. When a developer writes a regression test for a new PR and the test fails, they investigate whether the fault lies in the code or in the test~\cite{luo2014empirical,vahabzadeh2015empirical}. LLM-based test generators instead treat the code as correct: they repair or discard failing tests~\cite{pizzorno2024coverup,gu2026testart}, so a bug in the PR is not surfaced.
This difference relates to the \emph{oracle problem}~\cite{barr2014oracle}. When the implementation is the main available description of behavior, LLMs may infer assertions from what the current code does, rather than what it is intended to do~\cite{konstantinou2024oracles}. When the code is defective, a generated test can thus encode the defect as expected behavior: the buggy implementation passes it, while a corrected version fails. We call such tests \emph{fault-enforcing} (Figure~\ref{fig:motivation}).

Fault-enforcing tests become particularly risky in evolving projects: They appear harmless at creation because all tests passes, so the developer gets no signal that anything is wrong, and the test enters the suite as if it were a validated expectation. Its cost appears later. When the fault is eventually found and fixed, the test fails on the correct code, and the developer has to decide whether the failure is a regression or a wrong test, which can require examining recent changes to both the production and the test code~\cite{vahabzadeh2015empirical}. Until then, the test makes the faulty behavior look specified: it adds coverage and test count without verifying intended behavior. In an evolving project that merges generated tests across many PRs, such tests can accumulate, each making an existing fault harder to fix and the suite less trustworthy over time.

This raises a fundamental question: 
\begin{quote}
\emph{Do LLM-generated regression tests enforce faults that persist and accumulate as the project evolves?}
\end{quote}
We study this question on three open-source GitHub projects, SciPy, Qiskit, and pandas. We inject faults into historical PRs and generate regression tests for each of them, simulating a developer who accepts the generated tests without validating them. Such behavior is common in a vibe-coding setting~\cite{fawzy2026vibe} where developers trust whatever the LLM produces. We then track whether the enforced faults persist through later project changes, accumulate across PRs, and escape detection by the project's test suite.

\begin{figure}[t]
\centering
\includegraphics[width=0.95\textwidth]{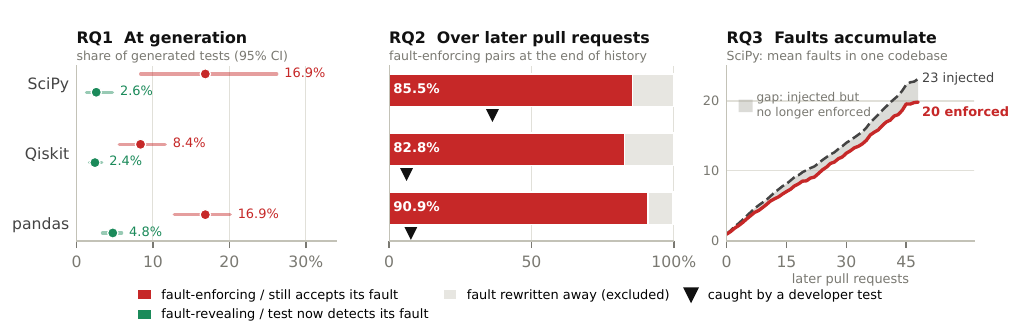}
\caption{When generating LLM-based regression tests for faulty PRs, 8.4\%--16.9\% of faults are enforced by the generated test suite, i.e. they pass on the faulty post-PR code and fail on the intended (left). Faults that are enforced often survive until the end of the studied project history (in 82.8\% -- 90.9\% of cases; middle). Faults accumulate (right; SciPy as an example): of 23 faults injected into a single codebase on average, 20 remain enforced at the end of our simulation.}
\label{fig:glance}
\end{figure}

Across 145 PRs (and 60,780 injected faults; Figure~\ref{fig:glance}), we find that LLM-generated regression tests enforce faults consistently more often than they reveal them. Depending on the project, 8\%--17\% of the generated tests enforce the injected fault, while only 2.4\%--4.8\% reveal it. Enforced faults are also durable: after several later commits, 83\%--91\% of fault-enforcing tests still pass only on the faulty code, so fixing the fault would still break them. They accumulate as well: when the faults of all PRs are combined in one codebase, 83\%--92\% remain enforced by a generated test at the end of the commit history, and the project's test suite detects only 14\%--30\% of them. Without manual validation, generated regression tests can therefore lock faults into the test suite.

Overall, our results shows that LLM-based regression testing can give rise to a new form of technical debt, in which accumulated fault-enforcing tests perpetuate defects and inhibit software evolution.

\inlineheadingbf{Contributions} This paper makes the following contributions:
\begin{itemize}
\item A \textbf{behavioral taxonomy} of LLM-generated regression tests with four categories (fault-revealing, fault-enforcing, weak, invalid), which classifies each test by its behavior on the faulty and the fixed code and thus measures test quality beyond coverage, together with a mutation-based methodology that measures it by injecting faults into historical PRs and tracking the generated tests through the commit history.
\item Two \textbf{longitudinal analyses} of fault-enforcing tests: \emph{persistence} of individual tests across later commits, and \emph{accumulation} of enforced faults when the faults of many PRs are combined in one codebase, including whether the project's test suite detects them.
\item An \textbf{empirical study} on 145 PRs from SciPy, Qiskit, and pandas. Fault-enforcing tests are 8\%--17\% of the generated tests, versus 2.4\%--4.8\% fault-revealing; 83\%--91\% of them persist; and when faults accumulate, 83\%--92\% remain enforced while the project's test suite detects only 14\%--30\% of them.
\end{itemize}

\section{Background and Related Work}
\label{sec:related}

\subsection{LLM-Based Regression Test Generation}

Automated test generation has traditionally targeted the generation of regression tests at unit- or class-level with EvoSuite~\cite{fraser2011evosuite} and Pynguin~\cite{lukasczyk2022pynguin} as popular examples. With recent advances in LLMs and the focus on more change-driven development, recent systems have started to focus more on generating tests at change-level, targeting lines modified in a commit or pull request.
ChaCo~\cite{zhou2026chaco} generates pull-request-aware tests that cover uncovered lines in merged patches; Liu~et~al.~\cite{liu2025regression} show that LLMs can produce commit-level tests that reproduce bug-introducing changes; Harman~et~al.~\cite{harman2025mutation} generate mutants for specific code-level issues at Meta and have an LLM write tests that kill them, evaluated by the mutants killed and by engineer acceptance; and Haratian~et~al.~\cite{haratian2026praware} find that pull-request-aware generation, by an LLM or by EvoSuite, rarely yields tests that capture the change. 

A second line of work steers generation towards coverage: SymPrompt~\cite{ryan2024symprompt} with code-aware prompts (five times as many correct tests with CodeGen2, more than twice the coverage with GPT-4), CoverUp~\cite{pizzorno2024coverup} with coverage feedback for Python suites, and TestForge~\cite{jain2025testforge} and LLMLOOP~\cite{ravi2025llmloop} with agentic refinement loops. Explicit test intentions~\cite{nan2025intention} and richer code context~\cite{walczak2025context} further improve generated tests.
These works evaluate generated tests by coverage, killed mutants, change capture, or engineer acceptance. To the best of our knowledge, none examines whether such tests enforce faults, or how they behave as the project evolves.

\subsection{Change-Aware Mutation Testing}

Change-aware mutation testing restricts mutation to code affected by recent changes. MuDelta~\cite{ma2021mudelta} introduces commit-relevant mutants, which correlate more strongly with real bugs than uniformly sampled ones; commit-aware mutation testing~\cite{ma2020commit} and studies across revisions~\cite{ojdanic2021mutation,ojdanic2022use,ojdanic2022change} report gains in efficiency and effectiveness. Recent work combines mutation with LLMs: \textmu{}BERT~\cite{degiovanni2022mubert} generates realistic mutants with a pre-trained language model, Straubinger~et~al.~\cite{straubinger2025mutation} use LLMs in a scientific-debugging loop to improve tests and mutation scores, and AdverTest~\cite{chang2026advertest} pits a test-writing agent against a mutant-writing agent to strengthen tests for a program assumed correct. Closest to our work, Haroon~et~al.~\cite{haroon2026evolution} evaluate LLM-based test generation on 22,374 program variants and find that many generated tests remain aligned with the original behavior under semantic-altering changes. Their changes are small mutations of seed programs, and tests are generated per variant, whereas we track generated tests through real PR history.
We use commit-aware mutants~\cite{ma2021mudelta,ma2020commit} as injected faults, a proxy for real bugs, and study the persistence and accumulation of fault-enforcing tests across PR history, which, to our knowledge, prior work has not studied.

\subsection{The Oracle Problem in LLM-Based Testing}

The effectiveness of a test depends on the correctness of its oracle. While generating executable test inputs has become increasingly practical, generating assertions that capture \emph{intended} rather than \emph{observed} behavior remains a fundamental challenge~\cite{barr2014oracle, molina2024oracle}. Classic generators such as EvoSuite~\cite{fraser2011evosuite} and Randoop use regression oracles that assert the current behavior and thus assume the code is correct~\cite{dinella2022toga}.

Recent work revisits the problem for LLMs. Konstantinou~et~al.~\cite{konstantinou2024oracles} empirically show that LLM-generated test oracles are prone to capture actual program behavior rather than intended behavior, although the generated tests reach higher fault-detection potential than EvoSuite on the studied benchmarks. A follow-up study~\cite{konstantinou2026coding} further shows that tests generated from faulty code inherit its faults, lowering fault detection from 25\% to about 14\% on function-level benchmarks. Molinelli~et~al.~\cite{molinelli2025oracles} conduct a large-scale empirical study showing that LLMs struggle to produce useful oracles for complex specifications despite syntactic correctness. Bodicoat~et~al.~\cite{bodicoat2026oracle} provide an empirical study of oracle generation under diverse program characteristics. TOGLL~\cite{hossain2025togll} explores strategies for generating stronger and more correct test oracles from LLMs.

These studies evaluate oracles at a single point in time, mostly on function-level benchmarks, whereas the generation systems above are typically judged by whether tests pass and cover the changed code. A test that encodes incorrect expectations passes on, and covers, the faulty code, so such evaluation rewards it. To our knowledge, none of these studies follows such tests through a project's history. Our empirical study captures this failure mode, enabling its systematic measurement and longitudinal evaluation.

\section{Study Design}
\label{sec:framework}
We investigate how LLM-generated regression tests behave as a project evolves with three research questions:
\begin{description}[style=unboxed, leftmargin=0pt]
\item[\textbf{RQ1.}] \emph{How often do LLM-generated regression tests enforce faults?}
\item[\textbf{RQ2.}] \emph{How often do fault-enforcing tests persist across later pull requests?}
\item[\textbf{RQ3.}] \emph{When faults accumulate in one codebase, how many remain enforced by generated tests, and does the project's test suite detect them?}
\end{description}

\begin{figure*}[t]
\centering
\includegraphics[width=0.9\textwidth]{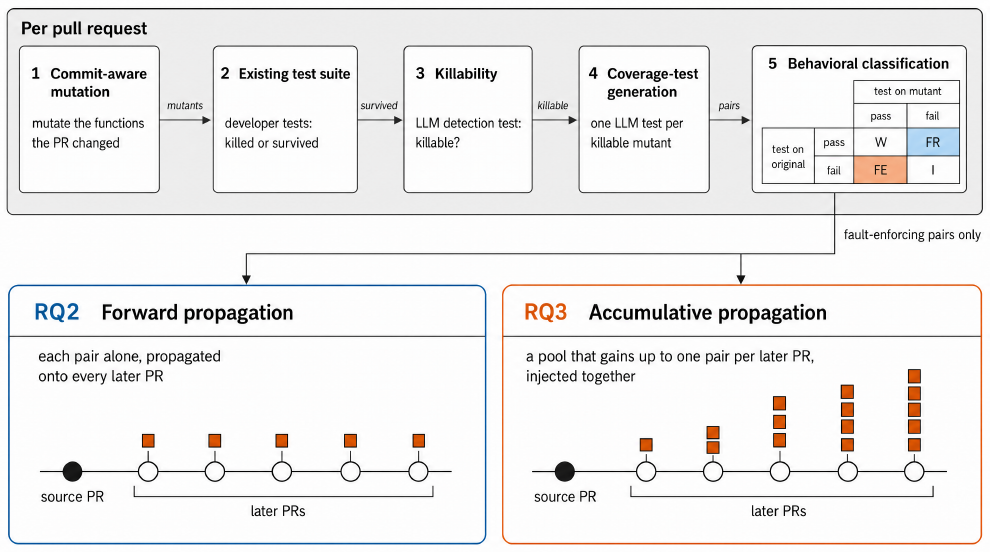}
\caption{\textbf{Empirical study.} Stages 1--5 run once per pull request and classify every test $T$, generated for a specific fault $C_\text{fault}$ (RQ1); the table in stage 5 defines the classes (Section~\ref{sec:taxonomy}). Only fault-enforcing pairs (squares) continue: forward propagation (RQ2) carries each through every later pull request (circles), and accumulative propagation (RQ3) pools one pair per later pull request and measures them together.}
\label{fig:pipeline}
\end{figure*}

To answer them, we operate at the level of pull requests. We treat a PR as a single change that transforms the code before the PR, $C_\text{old}$, into the changed code $C_\text{new}$\footnote{For a PR with several commits, we consider the merged state after the PR as $C_\text{new}$.}. We take $C_\text{ref} := C_\text{new}$ as the reference (fixed) version and obtain faulty versions $C_\text{fault}$ as mutants of $C_\text{new}$ that survive the project's test suite but can be killed by some test (Section \ref{sec:injection}). A faulty PR would therefore pass the project's own tests and be merged. Regression tests are generated from the change between $C_\text{old}$ and $C_\text{fault}$, without access to $C_\text{ref}$, simulating a developer who submits a faulty PR. We run each generated test on $C_\text{fault}$ and $C_\text{ref}$ and classify it by the outcome (Section \ref{sec:taxonomy}). 
RQ2 and RQ3 then follow the faults and their tests through the later pull requests (Section~\ref{sec:propagation}). Figure~\ref{fig:pipeline} provides an overview.

\subsection{Behavioral Taxonomy}
\label{sec:taxonomy}

Let $\mathcal{T}$ be the set of regression tests generated from the change between $C_\text{old}$ and $C_\text{fault}$. We run each test $T \in \mathcal{T}$ on $C_\text{fault}$ and $C_\text{ref}$. Based on its two outcomes, we assign $T$ to exactly one of four categories:

\begin{itemize}

\item \textbf{Fault-Revealing (FR):} $T$ fails on $C_\text{fault}$ and passes on $C_\text{ref}$, so it detects the injected fault.

\item \textbf{Fault-Enforcing (FE):} $T$ passes on $C_\text{fault}$ and fails on $C_\text{ref}$, so it asserts behavior of the faulty code that the reference does not exhibit.

\item \textbf{Weak (W):} $T$ passes on both. It may add coverage but does not distinguish the faulty and the reference behavior.

\item \textbf{Invalid (I):} $T$ fails on both, e.g., due to incorrect assertions or environmental assumptions that do not hold.
\end{itemize}

A fault-enforcing test \emph{silently accepts the fault}: it passes on the faulty code and thus gives the developer no signal of the fault, but it fails once the fault is fixed, which makes it a false alarm.

\subsection{Fault Injection}
\label{sec:injection}

We inject faults by mutating $C_{new}$. Each mutant pass the project's own test suite, so that the faulty PR would pass and be merged. We obtain them in three stages.

\inlineheadingbf{Stage 1: Commit-Level Mutation}
For each pull request, we apply standard mutation operators to the functions that the PR modifies and, when a method of a class changes, to the other methods of that class. This scopes mutation to the neighborhood of the change, following the commit-aware paradigm~\cite{ma2021mudelta,ma2020commit}. 

\inlineheadingbf{Stage 2: Test Survival} We assume that a faulty PR whose fault is detected by the existing test suite would not be merged and would trigger no further testing effort. We therefore run the project's test suite after applying the PR on every mutant, considering only tests that pass on $C_\text{new}$. A failure of a test that already fails on $C_\text{new}$ says nothing about the mutant. If some considered test fails on a mutant, the suite detects the fault and we discard the mutant. Only surviving mutants continue to the next stage; they represent faulty PRs that pass the project's tests and are merged, which is where additional regression tests can make a difference.

\inlineheadingbf{Stage 3: Killability} A surviving mutant may be \emph{equivalent} to $C_\text{new}$, i.e., behave identically on all inputs, or differ only in ways that no reasonable test can observe. Such a mutant is not a fault, and every test for it ends up weak or invalid, which would distort our results. Since equivalence is undecidable in general, we approximate it empirically. For each surviving mutant, an LLM receives the original and the mutated function and writes a \emph{detection test}. A mutant is \emph{empirically killable} if, within $k$ attempts, at least one such test passes on $C_\text{new}$ and fails on the mutant. We keep only killable mutants, so that each $C_\text{fault}$ contains a fault that the existing suite misses and that some test can detect.

\subsection{Test generation}
\label{sec:generation}

Given a faulty version $C_\text{fault}$, we simulate a developer who generates regression tests for their PR with an LLM (Stage~4 in Figure~\ref{fig:pipeline}). We follow the design shared by recent coverage-guided, pull-request-aware approaches~\cite{zhou2026chaco,pizzorno2024coverup,jain2025testforge,ravi2025llmloop}: the model receives the change, generates a test that targets the changed lines, and refines it from execution feedback. Section~\ref{sec:setup} describes our implementation.

\inlineheadingbf{Input} For each killable mutant, the LLM receives the pre-commit code $C_\text{old}$, the mutated post-commit code $C_\text{fault}$, and the diff between them. It generates a \emph{coverage test} that targets the lines changed by the PR. It never sees $C_\text{ref}$.

\inlineheadingbf{Refinement} The model refines the test from execution feedback for up to six turns, until the test passes on $C_\text{fault}$ and covers the target lines. If the budget runs out, we keep the last attempt that covers the target lines.

\inlineheadingbf{Acceptance Criterion} We keep the acceptance criterion of existing regression test approaches unchanged: a test is accepted only if it passes on the code the generator sees. \emph{This is the risk we study.} An accepted test passes on $C_\text{fault}$, so it can only be weak or fault-enforcing by design. Therefore, we also keep tests that still fail on $C_\text{fault}$ when the refinement budget runs out. Those tests would have been discarded by existing approaches. Kept tests that also pass on $C_\text{ref}$ are fault-revealing or invalid otherwise.

\subsection{Longitudinal Analysis}
\label{sec:propagation}

Having identified fault-enforcing tests (Stage~5), we study what happens to them as the project evolves. We ask whether they persist across later changes (RQ2) and whether they accumulate when the faults of many PRs are combined in one codebase (RQ3).

\subsubsection{Persistence}

Let $PR_1, \dots, PR_n$ be the pull requests recorded in the project's history, and let $PR_i$ be the pull request into which we injected the fault. For every later $PR_j$ with $j > i$, we \emph{propagate} the fault: we apply the same mutation to $C_j$ (the code after applying $PR_j$) to obtain a faulty version $C^\text{fault}_j$~\cite{ojdanic2021mutation,falleri2014gumtree}. Together with it, we propagate each fault-enforcing test $T \in \mathcal{T}$ generated for $PR_i$. At $PR_j$, we classify $T$ with the taxonomy of Section~\ref{sec:taxonomy}, taking $C^\text{fault}_j$ as the faulty and $C_j$ as the reference version. This yields one of the following outcomes:

\begin{itemize}
\item \textbf{Excluded.} The mutation can no longer be applied to $C_j$, because the code it changed was rewritten or removed. There is no verdict: the fault has disappeared, but nothing detected it, and the test that accepted it remains in the suite.
\item \textbf{Survived} ($T$ is fault-enforcing). The test still accepts the fault.
\item \textbf{Killed} ($T$ is fault-revealing). The test now catches the fault, in the ordinary mutation-testing sense.
\item \textbf{Weakened} ($T$ is weak). The test no longer distinguishes the fault.
\item \textbf{Broken} ($T$ is invalid). The test fails on both versions, e.g., due to changes in the environment or the code under test.
\end{itemize}
All outcomes except Survived are terminal: $T$ leaves the analysis at that PR.

\inlineheadingbf{Baseline gating} Between pull requests, the surrounding codebase drifts: imports move, dependencies change, and a test can fail for reasons unrelated to any mutation. Every claim we are making at a pull request level is therefore gated based on a clean baseline: the generated test is first run on the \emph{original} codebase after the reference pull request is merged. A run that ends in a collection error or a timeout is skipped. We also pre-filter developer tests: Before being considered to kill a propagated fault, it must first pass on the original codebase.

\subsubsection{Accumulation}

Persistence follows one fault at a time, but faults can also accumulate: a project that merges fault-enforcing tests at successive changes may end up with several latent faults at once. We simulate such histories. Starting from a pull request $PR_i$, we step through the later PRs and at each $PR_k$ add the fault of one randomly chosen fault-enforcing test generated for $PR_k$ (at most one per source file), together with that test. At every $PR_k$, all faults added so far are injected into the code at once, and each test runs against its own fault in the presence of the others. At the end of the history we measure how many tests still enforce their fault and whether the project's test suite detects the accumulated faults. We repeat the simulation N times per source PR with different random choices and report averages.

\section{Experimental Setup}
\label{sec:setup}

\subsection{Subject Systems}
We evaluate on SciPy~\cite{scipy}, a mature scientific-computing library with an extensive regression suite across numerical domains; Qiskit~\cite{qiskit}, IBM's quantum-computing framework, with a substantially different code and test structure; and pandas~\cite{mckinney2010pandas}, a widely used data-analysis library. We study pull requests that are merged throughout the project's history and generate regression tests for lines \emph{uncovered} by the existing test suite. We only consider merged pull requests that change code beyond deletions and documentation, and exclude backports, documentation-titled changes, changes outside the instrumented source tree (for pandas, its I/O module), and changes to more than five code files, which are mostly refactorings. To evaluate the impact of LLM-based regression tests, we focus on patches which change is not fully covered. This selection results in 121 SciPy, 135 Qiskit and 45 pandas pull requests. For our evaluation, we use all \textbf{45} pandas pull requests and randomly sample \textbf{50} pull requests from SciPy and Qiskit.
This results in 145 pull requests in total that we mutate to introduce potential faults in the implementation.

Each pull request provides the pre- and post-commit revisions, the diff, the developer-written tests, and a container image of the project at that pull request, which lets a mutant from an earlier pull request run against a later one. All five stages are completed for every pull request.

\subsection{Mutation Analysis}

Mutation analysis is restricted to functions modified by each pull request, following recent commit-aware mutation testing practice~\cite{ma2021mudelta,ojdanic2021mutation}, extended to sibling methods of a changed class as described in Section~\ref{sec:framework}. Standard mutation operators (arithmetic operator replacement, relational operator replacement, logical connector replacement, constant replacement, and return value modification) are applied to all selected functions.

This process generated \textbf{21,240 mutants} for SciPy, \textbf{19,534} for Qiskit and \textbf{20,006} for pandas, 60,780 in total. For each mutant, we run the tests of project's test suite that exercise the mutated function are executed. Mutants killed by the tests are excluded from our study. A mutant is also excluded if the test run ends in a collection error, a timeout, or if no relevant test file exists, because the result then says nothing about the mutant. We check the surviving mutants for killability (Stage~3) and keep only those for which an LLM writes, within up to 8 attempts, a test that passes $C_\text{new}$ and fails on the mutant.
Table~\ref{tab:overview} reports these counts.

\subsection{Coverage-Guided Test Generation}
We adopt a pull-request-aware prompting strategy, inspired by ChaCo~\cite{zhou2026chaco}, using Claude Haiku 4.5 (\texttt{claude-haiku-4-5-20251001}) to generate tests that cover the code change. The coverage-test are generated based on the modified source code $C_\text{fault}$, surrounding context, the diff from the pre-commit code $C_\text{old}$, and project-specific execution information. Generation is iterative: the model observes the outcome of running its previous attempt and may revise it, up to 6 turns, for a coverage test. The full parameters (prompts, token limits, acceptance rules, feedback formats) are in the replication package (Section~\ref{sec:data}).

\subsection{Forward Propagation}
\label{sec:provenance}

\begin{table}[tb]
\centering
\footnotesize
\setlength{\tabcolsep}{3pt}
\caption{Overview of Generated Mutant-Test Pairs}
\label{tab:flow}
\begin{tabularx}{\columnwidth}{@{}Xrrr@{}}
\toprule
\textbf{Stage} & \textbf{SciPy} & \textbf{Qiskit} & \textbf{pandas} \\
\midrule
Mutants generated & 21,240 & 19,534 & 20,006 \\
Stage~3 & 3,483 & 3,446 & 10,609 \\
Stages~4--5$^{a}$ & 3,483 & 3,355 & 9,578 \\
Fault-enforcing mutant-test pairs & 587 & 281 & 1,614 \\
Propagated mutant-test pairs$^{b}$ & 587 & 279 & 1,614 \\
\bottomrule
\end{tabularx}
\par
\begin{minipage}{\columnwidth}\footnotesize $^{a}$Killable mutants for which Stage~4 kept a test. $^{b}$The last pull request of each project has no later one and is never a source.\end{minipage}
\end{table}

To answer RQ2 and RQ3, we propagate mutants together with fault-enforcing test cases through the commit history of the project. Table~\ref{tab:flow} shows the number of mutant-test pairs that can be propagated to the last pull request in our selection.

\section{Results}
\label{sec:results}

\subsection{RQ1: Behavioral Quality of Generated Tests}
\label{sec:rq1}

\begin{table}[tb]
\centering
\footnotesize
\setlength{\tabcolsep}{3pt}
\caption{Mutation Analysis}
\label{tab:overview}
\begin{tabularx}{\columnwidth}{@{}Xrrr@{}}
\toprule
\textbf{Metric} & \textbf{SciPy} & \textbf{Qiskit} & \textbf{pandas} \\
\midrule
Pull requests               & 50              & 50              & 45 \\
Mutants generated           & 21,240          & 19,534          & 20,006 \\
Not testable$^{a}$          & 696             & 1,384           & 2,535 \\
Mutants tested              & 20,544          & 18,150          & 17,471 \\
Killed by the existing suite   & 12,032 (58.6\%) &  9,806 (54.0\%) &  2,792 (16.0\%) \\
Survived the existing suite    &  8,512 (41.4\%) &  8,344 (46.0\%) & 14,679 (84.0\%) \\
\midrule
Killable: an LLM-written test detects it$^{b}$ & 3,483 (40.9\%) & 3,446 (41.3\%) & 10,609 (72.3\%) \\
\bottomrule
\end{tabularx}
\par
\begin{minipage}{\columnwidth}\footnotesize $^{a}$Collection error, timeout, or no relevant test file; excluded from all percentages. $^{b}$Percentage of survived mutants.\end{minipage}
\end{table}
Table~\ref{tab:overview} summarizes the mutation analysis across all three projects.
Out of 60,780 mutants, 56,165 mutants can be executed by the project's test suite without running into collection errors or timeouts. SciPy's and Qiskit's suites kill a majority of mutants in changed code (58.6\% and 54.0\%); pandas' kills 16.0\%. Of the survivors, 41--72\% are empirically killable; the rest are equivalent or hard to kill within the attempt budget. The budget is enough to identify most empirically killable mutants: we find that the eighth turn adds only 1.3\%, 4.3\% and 2.1\% of the killable mutants, indicating the effectiveness of the LLM converges.

\subsubsection{Behavioral Classification}

Table~\ref{tab:classification} gives the distribution of mutant-test pairs
for mutants that survived the developer suite, proved killable and received a coverage test. Three observations can be made. 

\begin{table}[tb]
\centering
\footnotesize
\setlength{\tabcolsep}{3pt}
\caption{Behavioral Distribution of LLM-Generated Regression Tests}
\label{tab:classification}
\begin{tabularx}{\columnwidth}{@{}Xrrr@{}}
\toprule
\textbf{Category} & \textbf{SciPy} & \textbf{Qiskit} & \textbf{pandas} \\
\midrule
Fault-Revealing (FR)   &    90 (2.6\%)  &    81 (2.4\%)  &   455 (4.8\%)  \\
Fault-Enforcing (FE)     &   587 (16.9\%) &   281 (8.4\%)  & 1,614 (16.9\%) \\
Weak (W)               & 2,251 (64.6\%) & 2,333 (69.5\%) & 3,961 (41.4\%) \\
Invalid (I)            &   555 (15.9\%) &   660 (19.7\%) & 3,548 (37.0\%) \\
\midrule
\textbf{Total}         & \textbf{3,483} & \textbf{3,355} & \textbf{9,578} \\
\bottomrule
\end{tabularx}
\par
\begin{minipage}{\columnwidth}\footnotesize 95\% intervals from a bootstrap that resamples whole pull requests: FR [1.3, 4.7], [1.7, 3.3], [3.4, 5.9]; FE [8.4, 26.2], [5.6, 11.6], [12.8, 20.1]; W [50.8, 77.9], [60.3, 77.0], [34.6, 46.4]; I [9.3, 24.8], [12.6, 28.5], [30.6, 46.3] (SciPy, Qiskit, pandas).\end{minipage}
\end{table}

\inlineheadingit{A substantial fraction are fault-enforcing} As expected, we find that a substantial fraction of LLM-generated regression tests enforce the fault. 16.9\%, 8.4\% and 16.9\% (587, 281 and 1,614) fail on the original yet pass on the mutant. The per-project shares are imprecise, SciPy's most (95\% interval 8.4--26.2\%), but the direction is not. In every project, the lower bound of the fault-enforcing interval lies above the upper bound of the fault-revealing one, with point ratios of 3.5--6.5$\times$. 

\inlineheadingit{Very few generated tests are fault-revealing} Surprisingly, even though automated regression testing is not designed for this purpose, we find that the LLM-generated regression tests are capable to reveal faults. In total, 2.6\%, 2.4\% and 4.8\% of pairs (90 of 3,483, 81 of 3,355 and 455 of 9,578) pass on the original implementation and fail on the mutant. These tests would have been discarded by a regular regression testing tool. Still, we have to note that the number remains substantial lower than for fault-enforcing tests: fault-enforcing pairs remain significantly more common than fault-revealing ones in every project (3.5--6.5$\times$; pull-request bootstrap). The gap holds for every mutation operator with at least 100 pairs, whichever operator is left out, and pull request by pull request (30 of 34, 26 of 32 and 28 of 33 pull requests where the shares differ; sign test and Wilcoxon signed-rank test, $p < 0.001$ in each project; matched-pairs rank-biserial 0.81--0.87).

\inlineheadingit{Most tests are still weak or invalid} Weak tests, e.g. those that are incapable to trigger the fault, dominate in SciPy (64.6\%) and Qiskit (69.5\%), while pandas splits between weak (41.4\%) and invalid (37.0\%). Invalid tests are often exercisable: only 249 of 4,763 (5.2\%) fail to import or collect; the rest runs and fails on the original and the faulty implementation with a regular test outcome.

\subsubsection{Per-Pull-Request Variance}
\label{sec:perpr}

\begin{figure}[tb]
\centering
\includegraphics[width=\textwidth]{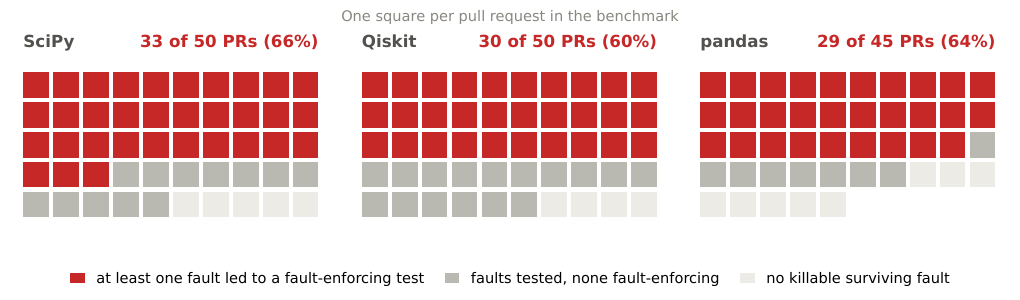}
\caption{Pull requests in which at least one injected fault led to a fault-enforcing test, one square per pull request of the benchmark. Grey: the pull request's faults were tested but none led to a fault-enforcing test; light: no surviving fault was killable, so no test was classified.}
\label{fig:perpr}
\end{figure}

\begin{figure*}[tb]
\centering
\includegraphics[width=\textwidth]{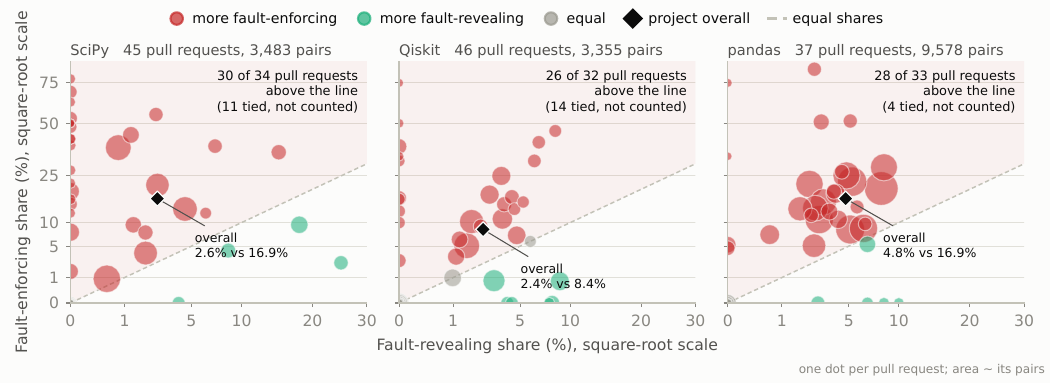}
\caption{Fault-revealing against fault-enforcing share, one dot per pull request (area proportional to its classified pairs), both axes on a square-root scale. The dashed diagonal marks equal shares; pull requests in the shaded area above it have more fault-enforcing than fault-revealing pairs. The diamond is the project's aggregate (Table~\ref{tab:classification}). Pull requests with equal shares, mostly none of either, sit on the diagonal and are not counted.}
\label{fig:perpr-scatter}
\end{figure*}

\begin{figure*}[tb]
\centering
\includegraphics[width=\textwidth]{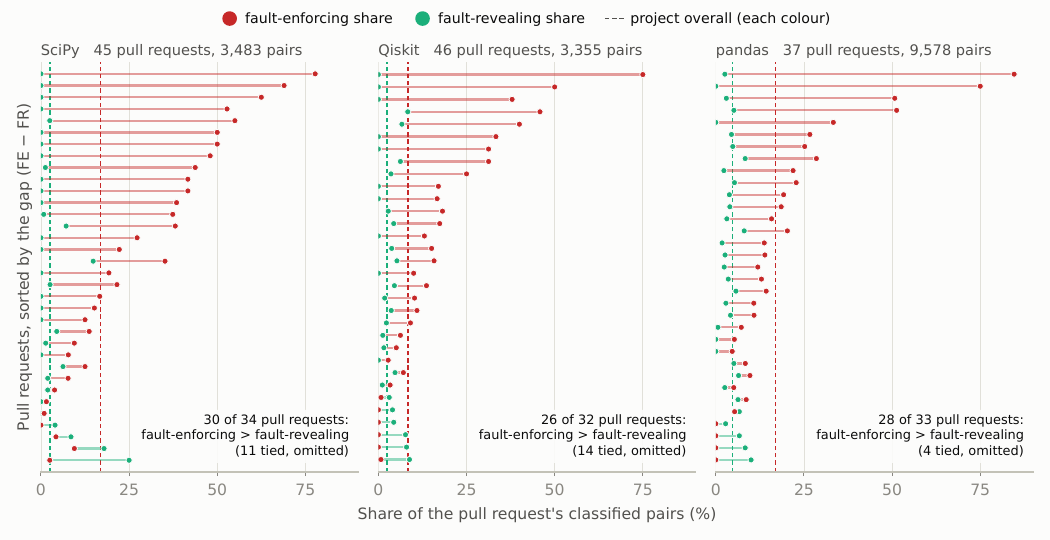}
\caption{Fault-enforcing (red) and fault-revealing (green) share of each pull request's classified pairs, one row per pull request, sorted by the difference between the two. The dashed lines mark the project's aggregate shares (Table~\ref{tab:classification}). Pull requests with equal shares, mostly none of either, are omitted.}
\label{fig:perpr-dumbbell}
\end{figure*}

Figure~\ref{fig:perpr} takes a worst-case view: in 92 of the 145 pull requests (63\%; 33 of 50 in SciPy, 30 of 50 in Qiskit and 29 of 45 in pandas), at least one injected fault led to a fault-enforcing test. In most pull requests, a single mistake by the developer can thus be locked in by a generated regression test. Figure~\ref{fig:perpr-scatter} compares the two shares within each pull request: a dot above the diagonal is a pull request with more fault-enforcing than fault-revealing pairs, which holds for 30 of 34 SciPy, 26 of 32 Qiskit and 28 of 33 pandas pull requests in which the two shares differ. Figure~\ref{fig:perpr-dumbbell} shows the same comparison one pull request per row, sorted by the difference between the two shares; the few pull requests with more fault-revealing pairs sit at the bottom.

\inlineheadingit{The spread is wide.} The highest fault-enforcing rates come from smaller pull requests (SciPy \#21505: 77.8\%, 9 pairs; Qiskit \#12979: 75.0\%, 4; pandas \#59114: 84.6\%, 39), while 12 of the 45 SciPy, 16 of the 46 Qiskit and 8 of the 37 pandas pull requests with classified pairs have none. The aggregate rate is thus driven by a subset of pull requests. However, the fault-enforcing share exceeds the fault-revealing share in every third of pull requests by size.

\subsubsection{What Fault-Enforcing Tests Assert}
\begin{figure}[t]
\centering
\includegraphics[width=\textwidth]{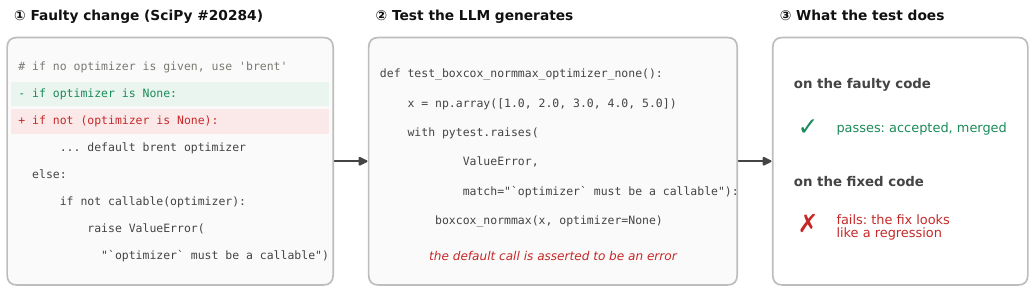}
\caption{A fault-enforcing test from SciPy \#20284 (test lightly shortened). The injected fault inverts the check for a missing optimizer; the generated test asserts that the default call raises an error, so it passes on the faulty code and fails once the fault is fixed.}
\label{fig:example}
\end{figure}

We categorized the assertions that enforce the fault. Exception expectations dominate with 37--57\% of enforcing tests expect an exception (e.g., \texttt{pytest.raises}), against 3--9\% of fault-revealing tests, which mostly check exact values, types or shapes. Exceptions are however not the cause for the higher number of fault-enforcing than fault-revealing tests: Even after filtering out all exception oracles, focusing on cases where the fault is directly encoded in test assertion, there are 4.0$\times$, 4.3$\times$ and 2.6$\times$ as many fault-enforcing tests as fault-revealing ones. 
Figure~\ref{fig:example} shows an example from SciPy \#20284.

\llbox{%
\textbf{Answer to RQ1.} Across all three projects, 8.4\% -- 16.9\% of all generated tests are fault-enforcing by enforcing the behavior of the mutant. 2.4--4.8\% are fault-revealing, while the remaining tests are either weak or invalid. Depending on the project, they share of fault-enforcing tests is several times higher than of the fault-revealing (3.5--6.5$\times$). In 99.8\% of comparable fault-enforcing pairs the accepted behavior did not exist before the pull request. 
}

\subsection{RQ2: Do Fault-Enforcing Tests Persist Through The Project's Evolution?}
\label{sec:rq2}

To answer RQ2, we follow each fault-enforcing test through the later pull requests of its project (Section~\ref{sec:propagation}). We assume that (1) a developer who does not detect the faulty PR merges the passing, fault-enforcing tests together with it, (2) when the mutated code is rewritten, the fault is no longer present and the pair leaves the analysis without a verdict, and (3) a developer test that fails on the mutant and passes on the original code would expose the fault. We do not stop tracking a pair when such a test exists, but report these cases separately. Table~\ref{tab:propagation} shows the results.

\begin{table}[tb]
\centering
\footnotesize
\setlength{\tabcolsep}{3pt}
\caption{Forward Propagation of Fault-Enforcing Pairs (RQ2)}
\label{tab:propagation}
\begin{tabularx}{\columnwidth}{@{}Xrrr@{}}
\toprule
\textbf{Final outcome of the pair} & \textbf{SciPy} & \textbf{Qiskit} & \textbf{pandas} \\
\midrule
Survived: a function still accepts the fault & 502 (85.5\%) & 231 (82.8\%) & 1,467 (90.9\%) \\
Killed: a function now detects the fault & 0 (0.0\%) & 0 (0.0\%) & 7 (0.4\%) \\
Weakened: nothing fails either way & 1 (0.2\%) & 0 (0.0\%) & 0 (0.0\%) \\
Excluded: mutated code rewritten$^{a}$ & 84 (14.3\%) & 48 (17.2\%) & 140 (8.7\%) \\
\midrule
\textbf{Pairs propagated}       & \textbf{587} & \textbf{279} & \textbf{1,614} \\
Source pull requests            & 33 / 33      & 29 / 29      & 29 / 29 \\
Evaluations (pair, later PR)    & 10,878       & 5,906        & 36,504 \\
\ \ skipped by the baseline gate & 620 (5.7\%) & 97 (1.6\%) & 30 (0.1\%) \\
\ \ skipped, every function dead & 177 (1.6\%) & 70 (1.2\%) & 263 (0.7\%) \\
\ \ set aside by the propagation check$^{c}$ & 436 & 160 & 4,320 \\
\midrule
Fault caught by a developer test$^{b}$ & 213 (36.3\%) & 17 (6.1\%) & 123 (7.6\%) \\
\bottomrule
\end{tabularx}
\par
\begin{minipage}{\columnwidth}\footnotesize $^{a}$No finding possible: the bug is gone, but nothing detected it. $^{b}$Separate observation: a developer test failed on the bug at $\geq 1$ later PR. $^{c}$Not counted in the evaluations above: the pair's history is cut at the first evaluation that placed the mutant in a different function than at its source (Section~\ref{sec:provenance}).\end{minipage}
\end{table}

\subsubsection{Most Fault-Enforcing Tests Persist}
In every project, most fault-enforcing tests survive until the last pull request: 85.5\% in SciPy, 82.8\% in Qiskit, and 90.9\% in pandas. Almost all of the remaining pairs are excluded because the mutated code was rewritten (8.7\%--17.2\%); among pairs whose mutation still applies, 99.5\%--100\% survive. Killed and weakened pairs are negligible (at most 0.4\%). Surviving does not mean the fault stays unnoticed, however: a developer test fails on the fault at some later PR for 36.3\% of SciPy pairs, but for only 6.1\% and 7.6\% in Qiskit and pandas, and the generated test keeps accepting the fault in the meantime.

This suggests a concrete risk of LLM-generated regression tests: a fault-enforcing test persists alongside its fault and may give a false sense of assurance that the faulty behavior is intended. The fault then becomes harder to recognize as a fault, and harder to fix once it is found, because the persisting test fails on the corrected code. In Qiskit and pandas, where developer tests fail on the fault for at most 7.6\% of the pairs, the project's own tests rarely contradict the generated test.

\subsubsection{How Long Do Fault-Enforcing Tests Persist?}
The shares above say where each pair ends up, but pairs are not observed for the same length of time. A pair from an early pull request is followed through up to 49 later pull requests, a pair from a late one through only a few, and a pair leaves the study early when its mutated code is rewritten. Simply counting how many pairs survive to the end would mix these long and short histories. We therefore use a Kaplan--Meier estimate~\cite{kaplan1958nonparametric}, the standard way to measure survival when subjects are followed for different lengths of time and some leave early. A pair counts for as long as we observe it; when its history ends, its fault is rewritten away, or its test stops distinguishing the fault, it stops counting, without being counted as killed. Only a kill ends a pair's survival. Table~\ref{tab:survival} gives, after 5 to 40 later pull requests, the estimated share of pairs whose generated test still accepts its fault, and in parentheses how many pairs were still observed at that point.

The answer is: practically indefinitely. In SciPy and Qiskit, no fault-enforcing test ever starts to detect its fault, over up to 49 later pull requests. In pandas, the estimate drops once, early, to 99.5\%: seven pairs are killed at the second or third later pull request (described below), and none afterwards. The picture is the same in calendar time. Pairs are followed for a median of four, six and nine months (up to 14 months), the estimates do not change when measured in days or when pairs followed for less than six months are left out, and none of the 392 pandas and 63 SciPy pairs followed for more than a year is killed after its first month.

The seven pandas kills share one mechanism, and it is not a test learning the intended behavior. In each, the enforcing tests are dead at the killing pull request, and the detection comes from a sibling function that was green on both sides at the source. Five pairs from \#59114 that replace a dependency name with an empty string are caught at \#59254, where the rewritten \texttt{get\_dependency\_info} imports each entry and raises on the empty name; one from \#59141 is caught there when \texttt{\_\_finalize\_\_} reaches an attribute the new code no longer sets; and one from \#58370, a bad \texttt{axis} argument, is caught at \#58791 once the surrounding code validates it. These crashes, exposed by incidental coverage, are the only detections in 53,288 evaluations.

\begin{table}[tb]
\centering
\footnotesize
\setlength{\tabcolsep}{2.5pt}
\caption{Survival of Fault-Enforcing Pairs Over Time (RQ2)}
\label{tab:survival}
\begin{tabularx}{\columnwidth}{@{}Xrrrrr@{}}
\toprule
\textbf{Later PRs since source} & \textbf{5} & \textbf{10} & \textbf{20} & \textbf{30} & \textbf{40} \\
\midrule
SciPy & 100.0 (509) & 100.0 (449) & 100.0 (193) & 100.0 (166) & 100.0 (86) \\
Qiskit & 100.0 (239) & 100.0 (194) & 100.0 (121) & 100.0 (97) & 100.0 (47) \\
pandas & 99.5 (1,342) & 99.5 (1,230) & 99.5 (911) & 99.5 (526) & 99.5 (259) \\
\bottomrule
\end{tabularx}
\par
\begin{minipage}{\columnwidth}\footnotesize Each cell: estimated share (\%) of pairs whose generated test still accepts its fault after that many later pull requests (Kaplan--Meier), with the number of pairs still observed at that point in parentheses. Pairs whose history ends, whose fault is rewritten away, or whose test stops distinguishing the fault stop counting without counting as killed. pandas' 95\% interval: [98.9, 99.8].\end{minipage}
\end{table}

\subsubsection{Developer Tests Catch a Minority}
We also check whether the project's own tests detect the propagated fault. A fault counts as caught if a developer test that exercises the mutated function, and that passes on the unmutated code, fails on the mutant at some later pull request. This holds for 36.3\% of the SciPy pairs, but for only 6.1\% of the Qiskit and 7.6\% of the pandas pairs. In these two projects, over 90\% of the faults that the generated tests accept are never caught by a developer test within the studied history. This check does not involve the generated tests, so a catch means that the project's tests would have exposed the fault even without them.

\llbox{%
\textbf{Answer to RQ2.} Fault-enforcing behavior persists. Across 2,480 propagated pairs, 83--91\% survive to the end of the studied history with the generated test still accepting the fault, 9--17\% lose the fault to a rewrite of the mutated code. A developer test catches the fault at some later pull request for only 6--36\% of pairs.
}

\subsection{RQ3: Do Faults Accumulate in a Single Codebase?}
\label{sec:rq3}

In RQ1 and RQ2, we considered one fault at a time: whether a generated test enforces it, and whether the test keeps enforcing it across later pull requests. A project may also merge several faulty PRs over time, so faults can accumulate in the same codebase. To answer RQ3, we simulate such histories (Section~\ref{sec:propagation}). Starting from a source pull request, we add at each later pull request one randomly chosen fault together with its fault-enforcing test to the codebase, and run each test against its own fault in the presence of all the others. Every source pull request with a validated pair seeds up to three trials from a seeded random generator; in total, 246 trials from 91 source pull requests complete. Our experiments for RQ3 are a deliberately adverse stress test and not a model of how real faults accumulate. Table~\ref{tab:accumulation} summarizes the results.

\begin{table}[tb]
\centering
\footnotesize
\setlength{\tabcolsep}{3pt}
\caption{Accumulative Propagation of Fault-Enforcing Pairs (RQ3)}
\label{tab:accumulation}
\begin{tabularx}{\columnwidth}{@{}Xrrr@{}}
\toprule
\textbf{Measure} & \textbf{SciPy} & \textbf{Qiskit} & \textbf{pandas} \\
\midrule
Source pull requests          & 33 / 33    & 29 / 29    & 29 / 29 \\
Trials                        & 85         & 77         & 84 \\
Mean chain length (PRs)       & 26.2       & 22.9       & 22.2 \\
Mean final pool size, injected / eligible & 13.2 / 12.1 & 10.5 / 9.4 & 10.5 / 8.7 \\
Pool members, all trials      & 1,385      & 1,210      & 995 \\
\midrule
Compound evaluations$^{a}$    & 17,982     & 13,777     & 13,215 \\
Pool-induced kills            & 0 (0.00\%) & 51 (0.37\%) & 4 (0.03\%) \\
Members ever killed in a pool & 0 (0.0\%)  & 6 (0.5\%)  & 2 (0.2\%) \\
Culprit named by ablation     & --         & 37 / 51    & 4 / 4 \\
\midrule
Members caught by a developer test$^{b}$ & 419 (30.3\%) & 167 (13.8\%) & 209 (21.0\%) \\
\bottomrule
\end{tabularx}
\par
\begin{minipage}{\columnwidth}\footnotesize $^{a}$One run of an eligible test against its own fault, with the whole pool present. $^{b}$Confirmed: the developer test also fails with that member's fault alone. Percentages are per compound evaluation for kills, per pool member for members.\end{minipage}
\end{table}

\subsubsection{Faults accumulate}

\begin{figure}[tb]
\centering
\includegraphics[width=3.3in]{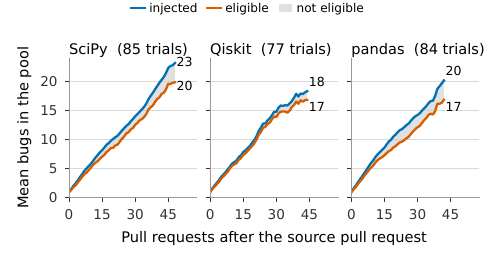}
\caption{Number of accumulated faults by number of pull requests since the introduction, averaged over trials. \emph{Injected}: faults present in the code at once. \emph{Eligible}: those counted in Table~\ref{tab:accumulation}, whose test still runs on the clean code of that pull request and still accepts its fault alone (fails without it, passes with it). The shaded gap: injected faults whose test no longer runs there (drift) or no longer accepts its fault alone.}
\label{fig:growth}
\end{figure}

Figure~\ref{fig:growth} shows how the pool of faults injected into a project evolve through the project history, averaged over the trials of a project. At position zero, the source pull request, the pool holds only the seed fault; each step is one later pull request at which it may gain one more. The upper line counts the \emph{injected} faults, those added so far minus the ones lost to attrition: the latent faults the codebase carries when the tests run. The lower line counts the \emph{eligible} members, those whose test still fails on the unmutated code and still accepts its own fault alone, the only ones for which a pool-induced kill is meaningful.

Growth is about one fault per two pull requests, since at many pull requests there is no fault and fault-enforcing test to add (616, 555 and 717 chain positions over all trials with no addition in SciPy, Qiskit and pandas) or every candidate would mutate a file the pool already mutates (170, 78 and 234). So, the longest chains reach 21--24 simultaneous bugs. Trials of one source pull request differ by about one member in final size. The eligible line trails the injected line by 8--17\% at the end of a chain, and kill rates in Table~\ref{tab:accumulation} are over eligible members only.

\subsubsection{Other Faults Rarely Change a Test's Behavior}

Across 44,974 compound evaluations, a function that accepted its fault alone failed with the rest of the pool present in only 55 cases. None in SciPy, 51 in Qiskit (0.37\%) and 4 in pandas (0.03\%); by member, 0, 6 and 2 (0.0\%, 0.5\% and 0.2\%) were ever killed in a pool. Where the ablation named a culprit (37 of Qiskit's 51 kills, all 4 of pandas'), it was always a mutant from a \emph{different} source pull request. The kills mostly come from a few trials as 44 of Qiskit's 51 come from two trials of one source pull request, and without its largest trial Qiskit's rate is 0.09\%; a trial-resampled 95\% interval for Qiskit's rate is [0.03, 0.99]\%. More bugs do not lead to more interaction: across all projects the rate is about 0.1\% for pools of 1--5, 6--10 and 11--15 simultaneous mutants (0.12\%, 0.12\%, 0.06\%) and 0.22\% above 15, where 19 of the 22 Qiskit kills come from the same source pull request; even the densest pools stay below 0.5\% at the upper bound of the 95\% interval.

The direction of this result matters as a pool-induced kill would mean that the incorrect expectation of a fault-enforcing test is fragile. We observe the opposite, a test that expects the faulty behavior keeps expecting it when the surrounding code carries a dozen other faults.

\subsubsection{The Project's Test Suite Detects a Minority}
We also run the project's own tests, those that exercise a mutated function and pass on the unmutated code, against the full pool of injected faults. A member counts as caught only if a test that fails with the pool also fails with that member's fault alone, so that a catch cannot be due to another fault. By this criterion, the project's tests catch 30.3\% of the SciPy, 13.8\% of the Qiskit, and 21.0\% of the pandas members. A further 374, 242, and 199 members (27.0\%, 20.0\%, and 20.0\%) fail a test only when the whole pool is present; these catches cannot be attributed to a member, so we do not count them. The confirmed rates are therefore lower bounds, and the project's tests could have caught at most 57\%, 34\%, and 41\% of the members. Percentages are over pool members across all trials, so a pair that appears in several trials is counted each time.
In Qiskit and pandas, the confirmed rates are higher than in RQ2 (6.1\% and 7.6\%), and in SciPy they are lower (36.3\%). The two measures differ in definition (confirmed against the fault alone versus any failure on the mutant) and in sampling (pool members across trials versus all pairs), so we do not compare them directly.

\llbox{%
\textbf{Answer to RQ3.} Faults and fault-enforcing tests accumulate over time. In 44,974 evaluations of a generated test on its own fault, in pools of up to 24 simultaneous faults, the other faults made the test stop accepting its fault in 0.00\%--0.37\% of cases, and at the end of a chain 83\%--92\% of the injected faults are still enforced by an eligible test. The project's test suite individually catches at least 14\%--30\% of the accumulated faults, and at most 34\%--57\%.
}

\section{Discussion}
\label{sec:discussion}

\subsection{Why Fault-Enforcing Tests Arise}
A fault-enforcing test specifies the behavior of the faulty code. Two factors produce it. The first is the oracle problem: a model that sees only the faulty implementation may take its behavior as the intended one and write an assertion accordingly. The second is the acceptance criterion commonly applied in LLM-based regression testing (Section~\ref{sec:generation}): a test is accepted only if it passes on the code the generator sees. An accepted test needs to pass on $C_\text{fault}$ and hence can only be weak or fault-enforcing, while a test that would reveal the fault is discarded or revised. We find that tests that would have been discarded when the refinement budget runs out can be fault-revealing. This design decision partly contributes to a large gap between 8\%--17\% fault-enforcing and 2.4\%--4.8\% fault-revealing tests.

\inlineheadingit{Regression Testing in general} Fault enforcement is a common problem of automated regression testing, and not specific to LLMs. Any oracle derived from observed behavior, whether written by an LLM or produced by a search-based generator such as Pynguin or EvoSuite, tends to encode what it observes. We additionally find that change-aware LLM generation, which also sees the diff and its context, does not avoid the problem, and that its consequences extend over time (RQ2 and RQ3). 

\subsection{Fault Enforcing Tests Do Not Vanish Over Time}
One might hope that in an ordinary project fault-enforcing tests are corrected through time: as developers extend the test suite and refactor code, incorrect assumptions would be eventually discovered or over-written. Our results however indicate that this is not the case for a large part of faults (and fault-enforcing tests): We find that 83\%--91\% of faults can still be applied and are accepted by the generated test. For 64\%--94\% of all faults no test of the project has failed on the fault by the end of the studied history.

\subsection{Implications}
\inlineheadingbf{For developers} A passing generated test is not evidence of correctness. Its expected values come from the code, so reviewers should check them against the specification or the PR description and not against the output of the current code.

\inlineheadingbf{For tool builders} Since a fault-enforcing test can only be recognized relative to a corrected version, intervention has to act on where the expectation comes from. Regression testers could derive expectations from specifications, documentation, or the PR description when these exist, following work on oracle generation from natural language~\cite{hossain2025togll}. Our study also indicates a {\em potential} of LLM-based (regression) testing: 2.4\% -- 4.8\% of tests are fault-revealing, even after several rounds of refinements. Therefore, tool builders could treat a failing test as a possible fault in the code, as developers do~\cite{luo2014empirical,vahabzadeh2015empirical}, and report it instead of repairing or discarding it. 

\inlineheadingbf{For evaluation} Coverage-guidance is a weak criterion: 41\%--70\% of the generated tests are weak, which means they cover the changed code but cannot tell apart the faulty version from the reference. Furthermore, evaluation at a single point in time cannot evaluate the impact of tests throughout the project evolution. We hence suggest to employ longitudinal benchmarks that track generated tests across pull request histories to give more realistic estimates of practical value.

\section{Threats to Validity}
\label{sec:threats}

\subsection{External Validity}
We evaluate with artificial faults (mutants), so our results may not transfer to real faults. Manual analysis of fault-revealing tests showed that many relate to obvious, unrealistic mutations. Simple mutants are easier to detect, and our killability filter keeps only faults that a model can detect, so the fault-revealing rate is likely optimistic. 

We further restrict ourselves to pull requests from three popular open source Python project. Although our results might not generalize to other unseen projects, we find that these projects span three major application domains in numerical computing, quantum computing and data manipulation, with our finding being consistent across all three domains. The projects are in addition all mature, widely used libraries with established regression suites, although their strength varies (the existing suites kill 58.6\%, 54.0\% and 16.0\% of testable mutants in changed code); projects with weaker testing cultures may differ.

We evaluate LLM-based regression testing with Claude Haiku 4.5, a model frequently used in practice and research for code and test generation. Our results might however not generalize to stronger models (that might have stronger capabilities to detect faults). However, we believe that our results are a fundamental consequence of the design of LLM-generated regression testing and the stronger capabilities of larger models to follow instructions might amplify our results. We leave this evaluation open for future research.

\subsection{Internal Validity}
Our taxonomy depends on correct mutation generation, execution, propagation and container management. All experiments ran in isolated per-pull request containers, categories are deterministic outcomes with no manual overriding, infrastructure errors are never read as results, and unit tests exercise every rule. Killability is relative to the model, so a mutant labeled as not killable within the budget may be equivalent or just beyond Claude Haiku 4.5. Drift is handled by the baseline gate and the per-test reading; a skipped evaluation (5.7\% at the gate for SciPy) cannot become a kill but lowers SciPy's effective sample. Forward propagation can misplace a mutant (Section~\ref{sec:provenance}); a misplaced mutant is far more likely to be skipped or killed than to incorrectly survive, so misplacement works against our continued survival finding. Restricting the longitudinal analyses to evaluations propagated without ambiguity leaves the survived share (85.4\%, 82.8\%, 90.4\%) and the interaction rates (0\%, 0.38\%, 0.03\%) nearly unchanged.

\subsection{Construct Validity}

Our framework assumes that mutation testing provides an appropriate proxy for measuring behavioral quality. Although mutants are a valid substitute for real faults in many settings~\cite{just2014mutants} and commit-relevant mutants relate to real faults~\cite{ma2021mudelta}, generated mutants do not perfectly represent all real programming defects. Standard mutation operators may overrepresent simple arithmetic or logical bugs and underrepresent algorithmic or data-structure defects common in numerical code. What we observe is that a test fails on the original post-commit implementation and passes on a mutant. So, that the test would enforce a real defect is an interpretation, which holds to the extent that the mutant resembles one.

The coverage test is generated from a diff between the \emph{pre}-commit source and the mutant, which spans both the pull request's change and the mutation. This alone cannot cause fault-enforcing behavior because the post-commit code and the mutant differ only by the mutation, so the mutation causes the outcome to change. However, the combined diff could make the model expect pre-commit behavior that the mutation happens to restore. We tested this by running every fault-enforcing test on the pre-commit code and found it in only 0.2\% of comparable pairs (Section~\ref{sec:rq1}).

Classification is by execution outcome, not assertion intent, so some weak tests may still document behavior. The one-mutant-per-file rule caps pool density, so RQ3's kill rate is a lower bound on denser accumulation, and it measures the stability of an incorrect expectation, not classical masking.

\subsection{Conclusion Validity}

Pull requests vary widely (Section~\ref{sec:perpr}), so we report intervals from bootstraps that resample whole pull requests and per-pull-request sign and Wilcoxon signed-rank tests. Every headline comparison holds at that level. Per-project point estimates are imprecise (SciPy's fault-enforcing share spans 8.4--26.2\%), so we claim directions and ranges rather than exact per-project rates. Pull requests with no propagated fault-enforcing pair (17 of 50 SciPy, 21 of 50 Qiskit and 16 of 45 pandas) are by construction absent from the longitudinal analyses, and all 91 eligible source pull requests have completed both analyses.

\section{Conclusion}
\label{sec:conclusion}

We presented an empirical study of whether LLM-generated regression tests enforce the behavior of (potentially faulty) PRs. To this end, we injected real-world PRs with artificial faults, generated regression tests for the faulty PRs, and then classified the test outcomes with a four-category taxonomy: fault-revealing, fault-enforcing, weak, and invalid. Our evaluation focused a coverage-guided, change-aware LLM-based approach that accepts only tests passing on the code it sees, and we evaluate on 145 pull requests from SciPy, Qiskit, and pandas. We find that 8\%--17\% of the generated tests are fault-enforcing and only 2.4\%--4.8\% are fault-revealing. Following 2,480 fault-enforcing pairs through later pull requests, 83\%--91\% still accept their fault at the end of the studied history, also where later changes touched the mutated function, and almost none comes to detect it. The project's test suite catches the fault for 6\%--8\% of the pairs in Qiskit and pandas and for 36\% in SciPy. In a deliberately adverse stress test with pools of up to 24 simultaneous faults, a test stopped accepting its fault in at most 0.4\% of evaluations.

This suggests a concrete risk of LLM-generated regression tests. \emph{When a faulty PR is accepted and merged, a generated test can become fault-enforcing: it passes silently on the faulty code, giving false confidence, and it fails once the fault is fixed, making the fix harder.}

Developers should therefore not treat a passing generated test as evidence of correctness: before merging, they should check its expected values against the intended behavior. Future work should derive expectations from specifications or PR descriptions instead of the observed code, and should treat a failing generated test as a possible fault in the code, not only in the test. It should also evaluate generation approaches that do not require tests to pass, to separate the effect of the acceptance criterion from the oracle problem.

\section{Data Availability}
\label{sec:data}

A replication package is available at \figsharelink. It contains the evaluation pipeline (mutant generation, test execution in per-pull request containers, behavioral classification, forward and accumulative propagation) with its tests, the benchmark configuration naming the 145 pull requests, the generated mutants and tests with every execution record, and the derived data and scripts behind every table and figure, which regenerate them in minutes without the raw data.

\bibliographystyle{ACM-Reference-Format}
\bibliography{references}

\end{document}